\documentclass[UKenglish,
    twocolumn,
    amsmath,
    amssymb,
    prl,
    aps,
    floatfix,
    superscriptaddress,
    10pt
]{revtex4-2}
\usepackage{float}
\usepackage{graphicx}
\usepackage{caption}
\usepackage{subcaption}
\usepackage{multirow}
\usepackage{tabularray}
\usepackage{amsmath}
\usepackage{amssymb}
\usepackage{mathtools}
\usepackage{dsfont}
\usepackage{amsthm}
\usepackage{mathrsfs}
\usepackage{physics}
\usepackage{braket}
\usepackage{qcircuit}
\usepackage{color}
\usepackage{xcolor}
\usepackage{ulem}
\usepackage{verbatim}
\usepackage{fancyhdr}
\usepackage{lipsum}
\usepackage{babel}
\usepackage[
    colorlinks=true,
    bookmarks=false,
    citecolor=blue,
    urlcolor=blue
]{hyperref}
\usepackage{orcidlink}
\hypersetup{colorlinks}
\newcommand{\ep}{\mathrm{e}}
\renewcommand{\j}{\mathrm{i}}
\newcommand{\ua}{\uparrow}
\newcommand{\da}{\downarrow}

\renewcommand{\d}{\mathrm{d}}

\renewcommand{\epsilon}{\varepsilon}

\newcommand{\cnn}{Universit\'e Paris-Saclay, Centre de Nanosciences et de Nanotechnologies, CNRS, 10 Boulevard Thomas Gobert, 91120, Palaiseau, France}
\newcommand{\quandela}{Quandela, 7 Rue Léonard de Vinci, 91300 Massy, France}
\newcommand{\cnnCite}{Universit\'e Paris Cité, Centre de Nanosciences et de Nanotechnologies, CNRS, 10 Boulevard Thomas Gobert, 91120, Palaiseau, France}

\newcommand{\nancy}{
Université de Lorraine, CNRS, LPCT, F-54000 Nancy, France}
\newcommand{\majulab}{MajuLab, CNRS-UCA-SU-NUS-NTU International Joint Research Laboratory, 117543 Singapore, Singapore}
\newcommand{\cqt}{Centre for Quantum Technologies, National University of Singapore, 117543 Singapore, Singapore.}

\begin{document}
\title{Accessing the dynamics of a single spin through the phase of a photonic qubit}

\author{Etienne Bargel\, \orcidlink{0009-0004-3245-4864}}
\email{etienne.bargel@universite-paris-saclay.fr} 
\affiliation{\cnn}
\affiliation{These authors contributed equally to this work.}

\author{Petr Steindl\,\orcidlink{0000-0001-9059-9202}}
\email{petr.steindl@universite-paris-saclay.fr}
\affiliation{\cnn}
\affiliation{These authors contributed equally to this work.}

\author{Adrià Medeiros \,\orcidlink{0000-0001-9059-9202}}
\affiliation{\cnn}

\author{Ilse Maillette de Buy Wenniger\,\orcidlink{0000-0002-8809-4704}}
\affiliation{Department of Physics, Imperial College London, London, UK}

\author{Dario A. Fioretto\,\orcidlink{0000-0003-3829-4000}}
\affiliation{\cnn}
\affiliation{\quandela}

\author{Stephen C. Wein\,\orcidlink{0000-0002-4332-4465}}
\affiliation{\quandela}

\author{Aristide Lema\^itre\,\orcidlink{0000-0003-1892-9726}}
\affiliation{\cnn}

\author{Isabelle Sagnes\,\orcidlink{0000-0001-8068-6599}}
\affiliation{\cnn}

\author{Maria Maffei\,\orcidlink{0000-0001-5183-4716}}
\affiliation{\nancy}

\author{Alexia Auffèves\,\orcidlink{0000-0003-4682-5684}}
\affiliation{\majulab}
\affiliation{\cqt}

\author{Olivier Krebs\,\orcidlink{0000-0002-9076-2708}}
\affiliation{\cnn}

\author{Lo\"{\i}c Lanco\,\orcidlink{0000-0003-3607-7539}}
\affiliation{\cnn}
\affiliation{\cnnCite}

\author{Pascale Senellart\,\orcidlink{0000-0002-8727-1086}}
\email{pascale.senellart@universite-paris-saclay.fr}
\affiliation{\cnn}

\preprint{APS/123-QED}

\date{\today}

\begin{abstract}

Spin-photon interfaces are key devices for quantum computing and communications. The spin acts as a quantum memory, optimally coupled to a photon that carries the same quantum information. Efficient spin-photon interfaces have been developed with semiconductor quantum dots in cavities, where the spin state information is mapped onto the polarization or time bin of a single photon. Here we report a new type of quantum dot-based spin-photon interface where the state of a single electron spin controls the phase of a single-rail photonic qubit. Measuring this optical phase through homodyne measurements reveals the dynamics of the spin precessing around a magnetic field and being subject to hyperfine-interaction decoherence. These observations open the way to a new type of spin-photon entanglement and new opportunities for applications in quantum technologies. 

\end{abstract}

\maketitle
\label{section_intro}

Spin–photon interfaces constitute fundamental building blocks for quantum communication and scalable photonic quantum computing~\cite{Wein_Ji_Wu_KimiaeeAsadi_Ghobadi_Simon_2020, Cirac_Zoller_Kimble_Mabuchi_1997, Beukers_Pasini_Choi_Englund_Hanson_Borregaard_2024, Borregaard_Pichler_Schröder_Lukin_Lodahl_Sørensen_2020, Economou_Lindner_Rudolph_2010, Lindner_Rudolph_2009}. They couple a stationary qubit, which can store and process quantum information locally, to a flying photonic qubit that can distribute this information. Long-lived spins, such as those of trapped atoms or solid-state colour centres, can support the distribution of entanglement over long distances \cite{Zhou_Malik_Fertig_Bock_Bauer_vanLeent_Zhang_Becher_Weinfurter_2024, Knaut_Suleymanzade_Wei_Assumpcao_Stas_Huan_Machielse_Knall_Sutula_Baranes_etal._2024, Krutyanskiy_Galli_Krcmarsky_Baier_Fioretto_Pu_Mazloom_Sekatski_Canteri_Teller_etal._2023, Humphreys_Kalb_Morits_Schouten_Vermeulen_Twitchen_Markham_Hanson_2018, Pompili_Hermans_Baier_Beukers_Humphreys_Schouten_Vermeulen_Tiggelman_dosSantosMartins_Dirkse_etal._2021, Rao_Yang_Wrachtrup_2015, Pieplow_Strocka_Isaza-Monsalve_Munns_Schröder_2023, Brekenfeld_Niemietz_Christesen_Rempe_2020, Babin_Stöhr_Morioka_Linkewitz_Steidl_Wörnle_Liu_Hesselmeier_Vorobyov_Denisenko_etal._2022}, while spin coherence times in the sub-microsecond range allow for local synchronization \cite{Bernien_Hensen_Pfaff_Koolstra_Blok_Robledo_Taminiau_Markham_Twitchen_Childress_etal._2013, Cabrillo_Cirac_García-Fernández_Zoller_1999, Delteil_Sun_Gao_Togan_Faelt_Imamoğlu_2016, Stockill_Stanley_Huthmacher_Clarke_Hugues_Miller_Matthiesen_LeGall_Atatüre_2017} and the deterministic generation of photonic graph states~\cite{Lindner_Rudolph_2009, Thomas_Ruscio_Morin_Rempe_2022, Tchebotareva_Hermans_Humphreys_Voigt_Harmsma_Cheng_Verlaan_Dijkhuizen_deJong_Dréau_etal._2019}.

Semiconductor quantum dots (QDs) embedded in optical cavities provide a particularly powerful implementation of such interfaces. The coupling of the QD to a confined optical mode has been engineered to allow the generation of single photons with record efficiency~\cite{Tomm_Javadi_Antoniadis_Najer_Löbl_Korsch_Schott_Valentin_Wieck_Ludwig_etal._2021, Ding_Guo_Xu_Liu_Zou_Zhao_Ge_Zhang_Liu_Wang_etal._2025, Huet_Lam_Pollet_Steindl_Bernard_Adiyatullin_Stepanov_Hease_Guilloux_Margaria_etal._2026}. Electrons and hole spins with coherence time up to the microsecond time scale have enabled spin-multi-photon entangement at high rates~\cite{Cogan_Su_Kenneth_Gershoni_2023,Coste_Fioretto_Belabas_Wein_Hilaire_Frantzeskakis_Gundin_Goes_Somaschi_Morassi_etal._2023, Huet_Ramesh_Wein_Coste_Hilaire_Somaschi_Morassi_Lemaître_Sagnes_Doty_etal._2025, Huet_Lam_Pollet_Steindl_Bernard_Adiyatullin_Stepanov_Hease_Guilloux_Margaria_etal._2026, Meng_Chan_Nielsen_Appel_Liu_Wang_Bart_Wieck_Ludwig_Midolo_etal._2024}. Recently, these progresses led to new,  hybrid spin–photon architectures in which local spin memories and photonic qubits are combined to reduce the resource requirements of fault-tolerant photonic quantum computing by several orders of magnitude \cite{Gliniasty_Hilaire_Emeriau_Wein_Salavrakos_Mansfield_2024, Dessertaine_Bourdoncle_Denys_Gliniasty_d’Istria_Valentí-Rojas_Mansfield_Hilaire_2026, Chan_Capatos_Lodahl_Sørensen_Paesani_2026, Chen_Löbl_Chan_Sørensen_Paesani_2026, Wein_GoubaultdeBrugière_Music_Senellart_Bourdoncle_Mansfield_2025, Hilaire_Vidro_Eisenberg_Economou_2023}.

Various encodings have been used to map the information from a QD spin state to a photonic state. Under low magnetic fields, the polarisation of the emitted photon is controlled by the spin state \cite{Lindner_Rudolph_2009, Coste_Fioretto_Belabas_Wein_Hilaire_Frantzeskakis_Gundin_Goes_Somaschi_Morassi_etal._2023, Köcher_Schumacher_2026}. Under high transverse magnetic fields, the optical transitions for the two spin states are spectrally distinct, so the spin state determines whether a photon is emitted under resonant excitation~\cite{Appel_Tiranov_Pabst_Chan_Starup_Wang_Midolo_Tiurev_Scholz_Wieck_etal._2022, Köcher_Schumacher_2026} resulting in an encoding based on the presence or absence of a photon. Both encodings have been used to demonstrate spin-photon entanglement and photonic cluster state generation using polarisation encoding~\cite{Lindner_Rudolph_2009,  Huet_Lam_Pollet_Steindl_Bernard_Adiyatullin_Stepanov_Hease_Guilloux_Margaria_etal._2026, Huet_Ramesh_Wein_Coste_Hilaire_Somaschi_Morassi_Lemaître_Sagnes_Doty_etal._2025} or time-bin encoding~\cite{Appel_Tiranov_Pabst_Chan_Starup_Wang_Midolo_Tiurev_Scholz_Wieck_etal._2022, Meng_Chan_Nielsen_Appel_Liu_Wang_Bart_Wieck_Ludwig_Midolo_etal._2024}. 
Here we report a new way to encode spin information onto a photonic wavepacket. We use a low magnetic field protocol and a coherent excitation of the QD. After erasing the polarisation information from the emitted quantum light, we obtain a quantum superposition of zero and one photon $\sqrt{1-p}\ket{0}+\ep^{\j(\varphi\pm\frac{\pi}{2})}\sqrt{p}\ket{1}$, that is, a single-rail qubit in which the spin state is mapped onto the $\pm\frac{\pi}{2}$ phase. This is evidenced experimentally by performing time-resolved self-homodyne measurements on the photonic qubit that reveal the spin dynamics. The photonic signal reflects the spin precession under the applied magnetic field as well as its decoherence. These observations constitute a first step toward a new spin-photon entanglement protocol and provide an estimate of the expected concurrence with the device under study.

\begin{figure}[!t]
\centering
\includegraphics[width=0.47\textwidth]{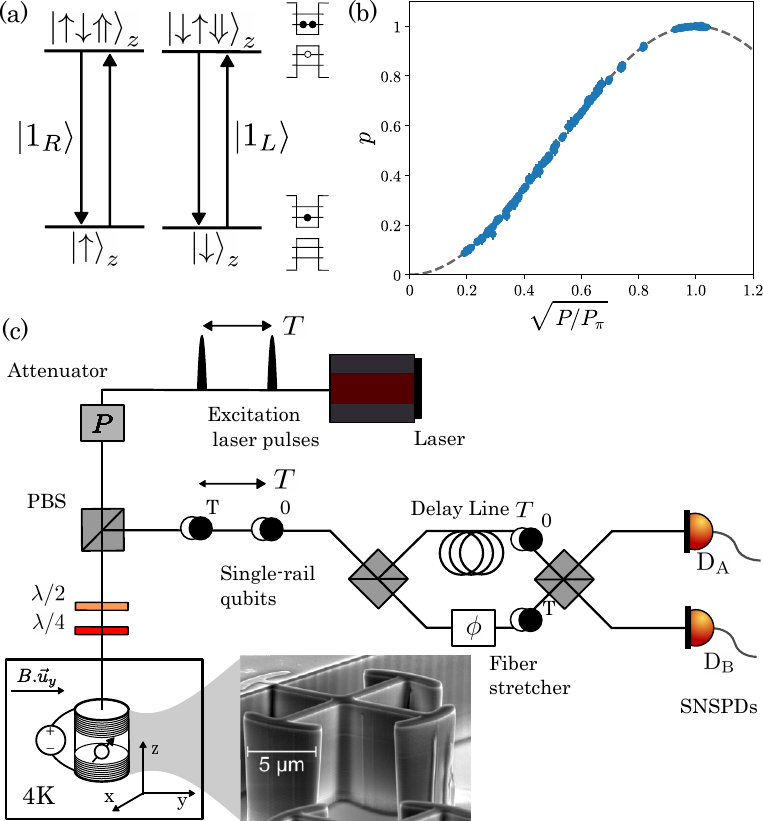}
\caption{\textbf{Spin-photon phase correlation measurement protocol.} (a) Optical selection rules of a negatively charged QDx. (b) Coherent control of the trion state population $p=\sin^2 \left(\Theta/2\right)$ using the excitation power $P$ of a resonant drive laser pulse, driving Rabi oscillations. (c) The experimental setup consists of a cross-polarized confocal microscope interfaced to a QD embedded in a micropillar cavity (SEM picture, inset) in a 4\,K cryostat~\cite{Steindl_Frey_Norman_Bowers_Bouwmeester_Löffler_2023} with an in-plane external electromagnet. The relative phase between two consecutive single-rail qubits extracted with relative delay $T$ is measured with a path-unbalanced Mach-Zehnder interferometer equipped with a pair of superconducting nanowire single-photon detectors (SNSPDs).
\label{fig_measurementprotocol}}
\end{figure}

{\it Spin-optical phase information transfer}.

The QD system considered here hosts a singly charged exciton, which, in the low magnetic field regime, can be described as a degenerate four-level system associated with the opposite spin projections of the resident carrier along quantization axis $z$. In the following, we focus on the negatively charged exciton, for which the electron spin ground states, $\ket{\uparrow}_z$ and $\ket{\downarrow}_z$, selectively couple to right- and left-circularly polarized optical transitions, denoted $\ket{1_R}$ and $\ket{1_L}$, respectively (Fig.~\ref{fig_measurementprotocol}\textbf{(a)}). The spin state is then mapped onto the polarization of the emitted photon~\cite{Acharya_Lanco_Krebs_Ng_Auffèves_Maffei_2026} according to
\begin{equation}
\left\{
\begin{array}{l}
\ket{\uparrow}_z \rightarrow \ket{\uparrow}_z\ket{1_R}, \\
\ket{\downarrow}_z \rightarrow \ket{\downarrow}_z\ket{1_L}.
\end{array}
\right.
\end{equation}
We additionally exploit the property demonstrated in 2019 whereby the coherent excitation of a two-level system enables deterministic generation of single-rail photonic qubits~\cite{Loredo_Antón_Reznychenko_Hilaire_Harouri_Millet_Ollivier_Somaschi_DeSantis_Lemaître_etal._2019}. For a generic emitter with ground and excited states $\ket{g}$ and $\ket{e}$, coherent excitation prepares the superposition state $\sqrt{1-p}\ket{g} + \ep^{\j\varphi}\sqrt{p}\ket{e}$ of population $p$ and optical phase $\varphi$ controlled by the excitation laser. Upon radiative decay, this coherence is transferred to the emitted field, yielding the single-rail qubit $\sqrt{1-p}\ket{0} +  \ep^{\j\varphi}\sqrt{p}\ket{1}$~\cite{Loredo_Antón_Reznychenko_Hilaire_Harouri_Millet_Ollivier_Somaschi_DeSantis_Lemaître_etal._2019}. Extending this picture to the two optical transitions of the charged exciton, an $H$-polarized coherent excitation generates single-rail qubits whose polarization are conditioned on the electron-spin state:

\begin{equation}
\left\{
\begin{array}{l}
\ket{\uparrow}_z \rightarrow
\ket{\uparrow}_z
\left(
\sqrt{1-p}\ket{0_R}
+
\ep^{\j\varphi}\sqrt{p}\ket{1_R}
\right), \\[0.4em]
\ket{\downarrow}_z \rightarrow
\ket{\downarrow}_z
\left(
\sqrt{1-p}\ket{0_L}
+
\ep^{\j\varphi}\sqrt{p}\ket{1_L}
\right).
\end{array}
\right.
\end{equation}
Considering an $H$-polarized coherent excitation that simultaneously drives the transitions $\ket{\uparrow}_z \rightarrow \ket{\uparrow\downarrow\Uparrow}_z$ and $\ket{\downarrow}_z \rightarrow \ket{\downarrow\uparrow\Downarrow}_z$ and collecting the emitted photons in the orthogonal polarization $V$ effectively transfers the polarization information $\ket{1_R/1_L} = \left(\ket{1_H}\mp\mathrm{i}\ket{1_V} \right)/\sqrt{2}$ onto the phase degree of freedom of the emitted single-rail qubit  $\ket{\varphi_p^\pm} \equiv  \sqrt{1-p}\ket{0_V} + \mathrm{e}^{\mathrm{i}(\varphi\pm\frac{\pi}{2})}\sqrt{p}\ket{1_V}$:
\begin{equation}
\left\{
\begin{array}{l}
\ket{\uparrow}_z
\rightarrow
\ket{\uparrow}_z\ket{\varphi_p^-},
\\[0.6em]
\ket{\downarrow}_z
\rightarrow
\ket{\downarrow}_z\ket{\varphi_p^+}.
\end{array}
\right.
\end{equation}
The two states $\left\{ \ket{\varphi_p^-}, \ket{\varphi_p^+}\right\}$ with opposite relative phases establish a direct information transfer between the spin state and the phase of the emitted single-rail qubit. We note that the same formalism applies to positively charged excitons upon a $\pi$ phase shift of the single-rail qubit. 

{\it Experimental demonstration.} We demonstrate such spin-phase correlations using an InGaAs quantum dot embedded in a GaAs/AlAs micropillar cavity exhibiting moderate mode splitting ($31.4~\mu \mathrm{eV}$). The QD is resonantly excited every $12.3\,\mathrm{ns}$ with $H$-polarized $10\,\mathrm{ps}$ laser pulses. The intensity of the collected $V$-polarized QD emission exhibits Rabi oscillations as the excitation-laser intensity $P$ is varied as shown in Fig.~\ref{fig_measurementprotocol}\textbf{(b)}. The pulse area, $\Theta=\pi\sqrt{P/P_\pi}$, controls the excited-state population $p=\sin^2(\Theta/2)$ demonstrating coherent control of the QD optical transition~\cite{Giesz_Somaschi_Hornecker_Grange_Reznychenko_DeSantis_Demory_Gomez_Sagnes_Lemaître_etal._2016}. For a laser intensity $P_\pi$ corresponding to a $\pi$-pulse resonant drive of the QD, deterministic generation of highly pure and indistinguishable single photons at high rates is achieved~\cite{Somaschi_Giesz_DeSantis_Loredo_Almeida_Hornecker_Portalupi_Grange_Antón_Demory_etal._2016}.  We apply a low in-plane Voigt magnetic field ($\le 60$\,mT) for which the optical selection rules presented above still apply, as experimentally evidenced by measuring a $M_s = (92.3\pm0.3)\%$ single-photon indistinguishability at $\pi$-pulse~\cite{Coste_Fioretto_Belabas_Wein_Hilaire_Frantzeskakis_Gundin_Goes_Somaschi_Morassi_etal._2023} for $B=60$\,mT. Single-rail qubits with population $p$ are generated for excitation powers $P<P_\pi$~\cite{Loredo_Antón_Reznychenko_Hilaire_Harouri_Millet_Ollivier_Somaschi_DeSantis_Lemaître_etal._2019}. 

\begin{figure}[!tb]
\centering
\includegraphics[width=0.47\textwidth]{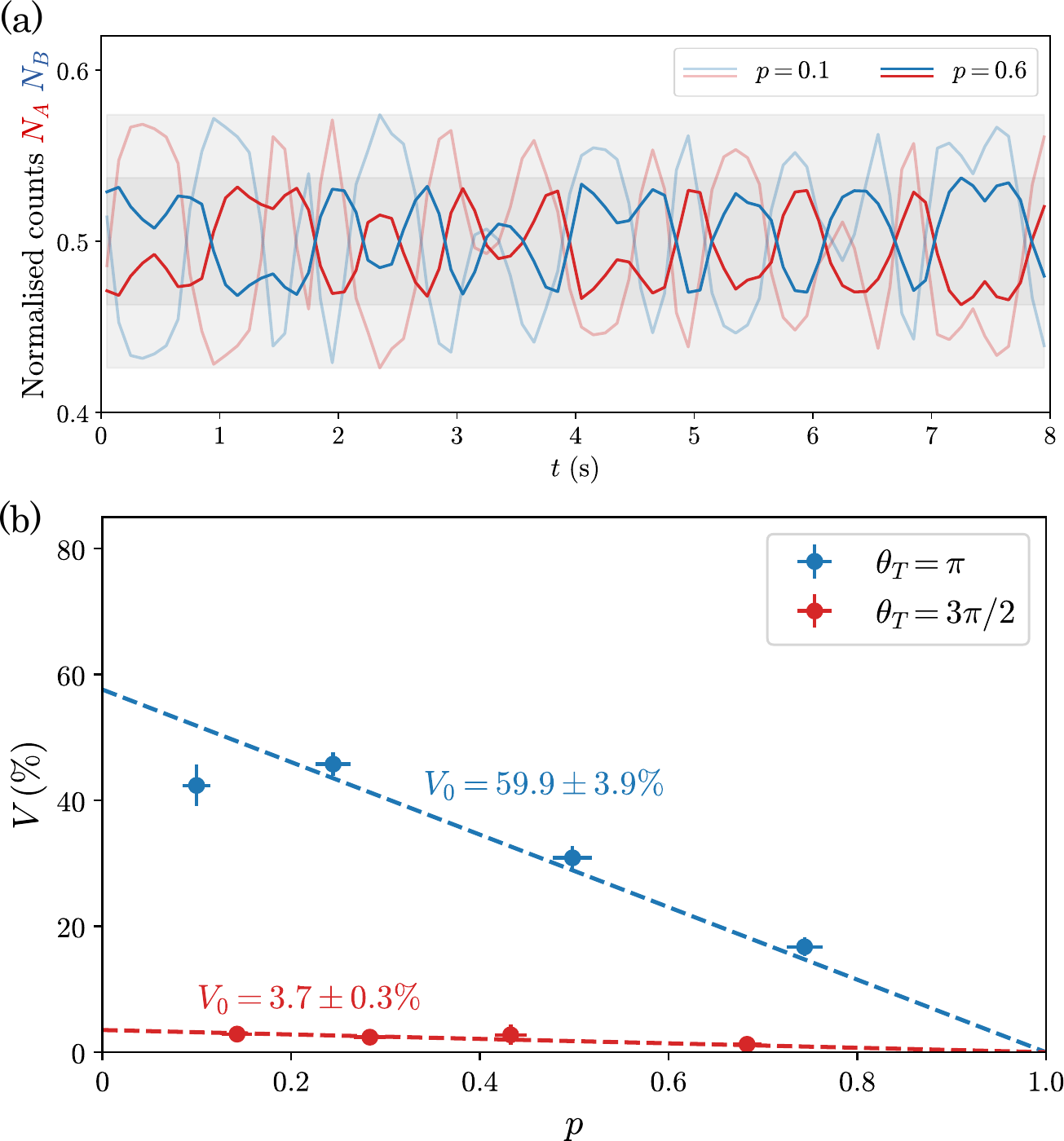}
\caption{\textbf{Spin-photon phase correlation measurement protocol.}
(a) Anti-correlated single-detector clicks registered at the output of the Mach-Zehnder interferometer as a function of the evolving internal phase $\phi(t)$ evidence the photon-number coherence. (b) The visibility of the oscillations scales with $1-p$, enabling extraction of $V_0$ by extrapolation to $p=0$, which value reflects spin dynamics during $T$, here given for two values of $\theta_T$.}
\label{fig_visibilityexample} 
\end{figure}

Spin-photon phase correlations are evidenced via single-rail qubit phase measurements based on the interference of these two single-rail qubits in a path-unbalanced Mach-Zehnder interferometer of internal delay $T$ in a self-homodyne measurement~\cite{Loredo_Antón_Reznychenko_Hilaire_Harouri_Millet_Ollivier_Somaschi_DeSantis_Lemaître_etal._2019}. As illustrated in Fig.~\ref{fig_measurementprotocol}\textbf{(c)}, two single-rail qubits are generated sequentially with identical populations, with a relative delay $T$, by two laser pulses. During this delay $T$, the electron spin precesses in the $x-z$ plane around the external Voigt magnetic field $B$ applied along the $y$ axis. This precession defines single-spin gates with an angle $\theta_T=\omega_LT$ at Larmor frequency $\omega_L = \left|g_e\right|\mu_B B/\hbar$, experimentally adjusted by varying the delay between the two laser probes and the amplitude of the applied magnetic field. Anti-correlated oscillations are observed in the normalized detected counts $N_A$ and $N_B$ at the interferometer outputs as shown in Fig.~\ref{fig_visibilityexample}\textbf{(a)}, evidencing the quantum coherence in the photon-number basis. These oscillations arise from interference between the two single-rail qubits and depend on the evolution of the varied internal Mach-Zehnder interferometer phase $\phi$ according to $N_{A/B} = p\left[1\pm(1-p)V_0\cos\phi\right]$. This enables the measurement of an interference visibility $V$ which accesses the magnitude of the normalized first-order coherence following $V = \underset{\phi}{\mathrm{max}}\frac{\left| N_A(\phi)-N_B(\phi)\right|}{N_A(\phi)+N_B(\phi)} =g^{(1)}(0,T)$. After renormalisation for the weight of the vacuum component of the single-rail qubit $1-p$, this enables extraction of the vacuum--one-photon interference visibility $V_0\equiv V/(1-p)$.

To understand how visibility measurement gives access to spin dynamics, we first consider an ideal non-decohering spin, initially in an arbitrary pure spin state, written in the $z$ basis as $\ket{\psi_s(0)} = \sqrt{p_\ua}\ket{\ua} +    \ep^{\j\chi}\sqrt{p_\da}\ket{\da}$ of respective spin state populations $p_\ua, p_\da$ such that $p_\ua+p_\da=1$, where $\chi$ is the relative phase between the two spin components. Under $H$-polarized excitation followed by emission at $t=0$ and cross-polarised $V$-collection, the spin state is encoded onto a spin-photonic state, with $\ket{\ua}_z$ and $\ket{\da}_z$ correlated with $\ket{\varphi_p^-}$ and $\ket{\varphi_p^+}$, respectively. During the delay $T$ between two excitation pulses, a magnetic field along $y$ rotates the spin by an angle $\theta_T$. This evolution is transfered onto the two consecutively emitted single-rail qubits and is captured by vacuum--one-photon interference visibility.

The phase sweep of the vacuum--one-photon interferometer gives access to the fringe visibility given by the spin-state trajectory on the Bloch sphere from $t=0$ to $T$: $V_0 = \sqrt{ \cos^2\theta_T + 4p_\ua p_\da\sin^2\theta_T\sin^2\chi }$. As expected, for $T=0$ no spin rotation is implemented ($\theta_T=0$) and maximal visibility is expected. Additionally, for a spin initially prepared along the rotation axis $y$ ($p_\ua=p_\da=1/2$ and $\chi=\pm\pi/2$), the absence of spin precession results in a time-independent visibility $V_0=1$ for any $\theta_T$. For a spin initialized in either $\ket{\ua}_z$ or $\ket{\da}_z$, , the interference visibility directly probes the spin rotation implemented by the gate, following $V_0 = |\cos\theta_T|$. This independence from the initial spin state within the $x-z$ plane also makes the relation valid for an initial incoherent mixture of $\ket{\ua}_z$ and $\ket{\da}_z$ at $t=0$, as encountered in our experiment where the spin is not initialized.

Indeed, in the presence of spin decoherence, the interference visibility measures the spin correlations between the two emission times (see Supplementary), without requiring prior spin initialization
\begin{equation} 
V_0 =\frac{|g^{(1)}(0,T)|}{1-p}=  \left|\langle \hat{S}_z(T)\hat{S}_z(0)\rangle\right|\,.
\end{equation}

\begin{figure}[!htbp]
\centering
\includegraphics[width=0.47\textwidth]{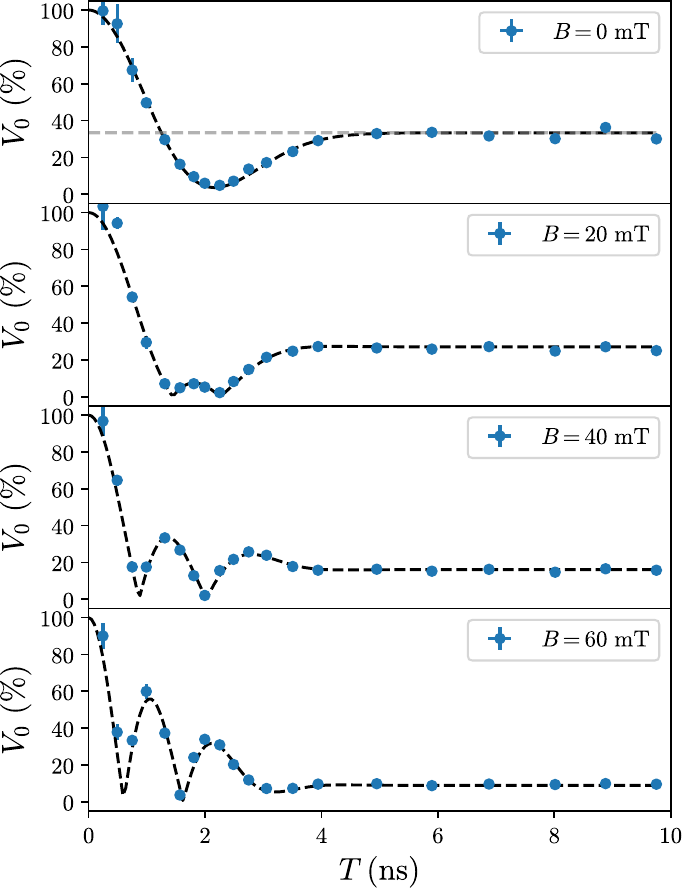}
\caption{\textbf{Spin-phase transfer in Merkulov-type measurement.} Vacuum--one-photon interference visibility $V_0$ between two photonic wavepackets coherently generated with varied temporal delay $T$. In different panels, the spin-state dynamics is varied by changing the in-plane magnetic field $B = 0$, $20$, $40$, and $60\,\mathrm{mT}$. Assuming initialized spin dynamics, the dashed lines represent theoretical spin evolution $\left|\langle \hat{S}_z (T) \hat{S}_z(0) \rangle\right|$ described by the Merkulov model~\cite{Merkulov_Efros_Rosen_2002}.
\label{fig_merkulov}}
\end{figure}

Figure~\ref{fig_visibilityexample}\textbf{(b)} presents the measured visibility $V$ for two different $\theta_T$, showing a linear dependence on $1-p$. A linear fit allows extrapolation to $p=0$ and extraction of visibility, $V_0$ with expected high (low) values for $\theta_T=\pi$ ($\theta_T=3\pi/2$). Figure~\ref{fig_merkulov} presents $V_0$ for various magnetic fields and time delays $T$, implementing different $\theta_T$. As expected, $V_0$ naturally approaches 1 at $T=0$ for all magnetic fields. When increasing $T$, various features appear depending on the amplitude of the magnetic field. To qualitatively understand these observations, and since the visibility is identical for a pure spin-up or spin-down state at $T=0$, we consider below an initial spin state $\ket{\ua}_z$.

For $B=0$\,mT, $V_0$ is found to decrease with the delay, reaching a minimal value characterized by the hyperfine-induced (or Overhauser) spin dephasing time $T_\mathrm{O}$ and then stabilizes toward $V_0\approx 1/3$ for long delays. This behavior faithfully reflects the well-known evolution of an electron spin in a bath of random nuclear spins without applied $B$~\cite{Urbaszek_Marie_Amand_Krebs_Voisin_Maletinsky_Högele_Imamoglu_2013, Merkulov_Efros_Rosen_2002}. During the first few nanoseconds, the dephasing is predominantly driven by the fluctuations of the Overhauser field with characteristic time scale $T_\mathrm{O}$. During this time, the initialized and freely evolving central spin is affected by stochastic spin flips in the nuclear bath, leading to its random reorientation and partial loss of its projection along its quantization axis $\langle \hat{S}_z \rangle$ \cite{Bechtold_Rauch_Li_Simmet_Ardelt_Regler_Müller_Sinitsyn_Finley_2015, Cogan_Kenneth_Lindner_Peniakov_Hopfmann_Dalacu_Poole_Hawrylak_Gershoni_2018}, captured by the Merkulov-Efros-Rosen model~\cite{Merkulov_Efros_Rosen_2002}, giving $T_O = 0.88\pm0.1\rm\,ns$. Without additional spin manipulation and on time scales beyond $T_\mathrm{O}$, the nuclear spin population is randomly oriented, leading to effective projections $\langle \hat{S}_z \rangle=1/3$.

When a magnetic field is applied, the visibility measurement reflects the spin precession evidenced by its oscillations damped by QD spin decoherence. 
We fit the measured $V_0$ by computing  $\left| \langle \hat{S}_z(T) \hat{S}_z(0) \rangle \right|$ assuming the electron spin dephasing induced by hyperfine interaction with a slowly varying nuclear bath~\cite{Merkulov_Efros_Rosen_2002}. Our simulations, presented as a broken line, show excellent agreement with the measured visibility. The corresponding parameters which describe the spin evolution are $T_O = 0.88\pm0.01\,\mathrm{ns}$ and the electron Lande-factor $\left|g_e\right| = \frac{\hbar}{\mu_BT_O\Delta_B} = 0.48 \pm 0.02$, from which we extract the amplitude of the effective magnetic field induced by nuclear spins $\Delta_B = 27.0 \pm 0.8\,\mathrm{mT}$ -- values consistent with those typically expected for similar negatively charged exciton devices and obtained from independent complementary measurement of $\left|g_e\right| = 0.5$ using magnetoluminescence~\cite{Medeiros_Gundín_Fioretto_Vinel_Rambeau_Mehdi_Somaschi_Lemaître_Sagnes_Belabas_etal._2026, Coste_Fioretto_Belabas_Wein_Hilaire_Frantzeskakis_Gundin_Goes_Somaschi_Morassi_etal._2023}. 

These observations provide clear evidence of the possibility of accessing the spin properties by measuring the phase of the single-rail qubits to which it is transfered under coherent excitation.

{\it Toward spin-phase entanglement. } Like for other encoding schemes, the present work can be extended to demonstrate entanglement between the spin state and the phase of the single-rail qubits. In the following, we restrict to balanced single-rail qubits $\ket{\varphi_{1/2}^\pm}$ with $p=\frac{1}{2}$, for which the two opposite-phase states are orthogonal and can therefore support maximal spin--photon phase entanglement.

Similarly to the polarization-encoded protocol ~\cite{Lindner_Rudolph_2009}, the sequence starts ($T=0$) with an initialized spin state, say in the $\ket{\ua}_z$ state. An applied transverse magnetic field along $y$ allows performing a $\pi/2$ spin gate for a given delay $T_{\pi/2}$ at which a single coherent excitation with pulse power $P_{\pi/2}$ is applied. Obtaining a maximally entangled state between the spin and the phase of single-rail qubits requires that the QD emits in a well-defined linear polarization. This can be achieved by coupling the QD into a strongly polarized cavity~\cite{Wang_He_Chung_Hu_Yu_Chen_Ding_Chen_Qin_Yang_etal._2019}. Owing to the polarization-selective Purcell enhancement given by such a cavity, the emitted field is preferentially extracted in the mode polarization upon radiative relaxation.
 
In the absence of decoherence, the obtained states $\frac{1}{\sqrt{2}} \left( \ket{\uparrow}_z\ket{\varphi_{1/2}^-} + \ket{\downarrow}_z\ket{\varphi_{1/2}^+} \right)$, shows maximal spin-photon phase entanglement, quantified by a concurrence $\mathcal{C}=1$. In the general case of an arbitrary $\theta_T$-spin rotation from $\ket{\uparrow}_z$, preparing $\cos(\frac{\theta_T}{2})\ket{\uparrow}_z + \sin(\frac{\theta_T}{2})\ket{\downarrow}_z$, spin-photonic state writes $\cos(\frac{\theta_T}{2})\ket{\uparrow}_z\ket{\varphi_{1/2}^-}+\sin(\frac{\theta_T}{2})\ket{\downarrow}_z\ket{\varphi_{1/2}^+}$ showing partial entanglement concurrence $ \mathcal{C}=|\sin(\theta_T)|$. In the presence of decoherence, the spin-photonic state density matrix in the $\left\{\ket{\ua}_z\ket{\varphi_{1/2}^-}, \ket{\ua}_z\ket{\varphi_{1/2}^+}, \ket{\da}_z\ket{\varphi_{1/2}^-}, \ket{\da}_z\ket{\varphi_{1/2}^+}\right\}$ basis is (see supplementary): 
\begin{equation}
\hat{\rho}_T = \frac{1}{2} 
\begin{pmatrix}
    1+\langle \hat{S}_z(T) \rangle & 0 & 0 & \langle \hat{S}_x (T) \rangle \\
    0 & 0 & 0 & 0\\
    0 & 0 & 0 & 0 \\
    \langle \hat{S}_x (T) \rangle & 0 & 0 & 1-\langle \hat{S}_z (T) \rangle
\end{pmatrix}\,.
\end{equation}

Interestingly, the concurrence of this entangled state is given by $\mathcal{C} = \left|\langle \hat{S}_x (T) \rangle\right|$, with $x$-axis spin projection~$\hat{S}_x$ (see Supplementary). The full characterization of the spin-photon entangled state would require a proper phase-spin homodyne tomography.  However, we note that the visibility measurement described above would directly give access to the entanglement concurrence through $\mathcal{C} = \sqrt{1-\left|\langle \hat{S}_z (T) \rangle\right|^2}=  \sqrt{1-V_0^2}$, starting from pure ground spin-state $\langle\hat{S}_z(0)\rangle=+1$. For the electron spin studied above, visibility measurements allow us to extract the parameters governing the spin dynamics within the decoherence model~\cite{Merkulov_Efros_Rosen_2002}, thereby reconstructing the spin evolution and predicting a maximum concurrence of $\mathcal{C}=65.9\,\%$ at $T=0.49\,\rm ns$ and $B=60\,\rm mT$ for a QD spin initialized along $z$ at $t=0$ (see Supplementary).

{\it Conclusions and perspectives. }
We have demonstrated a spin--photon interface in which the state of a single electron spin in a quantum dot is transferred onto the phase of a single-rail qubit. By measuring the phase of the emitted vacuum--one-photon state in a self-homodyne interferometer, we directly access the Larmor precession and hyperfine-induced decoherence of the spin. Complementary Hanle-type measurements~\cite{Bracker_Stinaff_Gammon_Ware_Tischler_Shabaev_Efros_Park_Gershoni_Korenev_etal._2005} performed at a fixed delay while varying the magnetic field amplitude (see Supplementary) provide an alternative way of tuning the accumulated Larmor precession and consistently reproduce the expected spin dynamics. We have further shown that this protocol can be extended to generate entanglement between the spin and the phase of a single-rail qubit. As for spin--photon interfaces based on polarization or time-bin encoding, we anticipate that present protocol can be extended to the sequential generation of spin--photon cluster states~\cite{Coste_Fioretto_Belabas_Wein_Hilaire_Frantzeskakis_Gundin_Goes_Somaschi_Morassi_etal._2023, Huet_Ramesh_Wein_Coste_Hilaire_Somaschi_Morassi_Lemaître_Sagnes_Doty_etal._2025, Thomas_Ruscio_Morin_Rempe_2022}, critical resources for fault-tolerant measurement-based quantum computing~\cite{Raussendorf_Briegel_2001,Paesani_Brown_2023,Raussendorf_Harrington_Goyal_2007}. 

The choice of the spin-photon interface encoding results in fundamental compromises on the quality of the spin-photon interface and possible applications~\cite{Wein_Ji_Wu_KimiaeeAsadi_Ghobadi_Simon_2020, Hermans_Pompili_SantosMartins_R-PMontblanch_Beukers_Baier_Borregaard_Hanson_2023}. For instance, time-bin protocols operate under resonant excitation, which guarantees an optimal single-photon emission efficiency, but put stringent requirements on the photon indistinguishability for computation since it requires a two-emission event~\cite{Gupta_Doherty_Mahmoodian_2026, Köcher_Schumacher_2026}. Polarisation-based protocols use non-resonant excitation, which can limit the single-photon emission efficiency, but are based on a single emission event, which supports computation protocols more tolerant on photon distinguishability~\cite{Gupta_Doherty_Mahmoodian_2026, Köcher_Schumacher_2026}. The present phase encoding offers a new route that relies on a low magnetic field and a single resonant coherent excitation, allowing for optimal single-photon emission. Moreover, contrary to polarization or time-bin encoding, single-rail phase encoding is naturally suited for the manipulation of the quantum information in photonic integrated chips, something that remains quite challenging with the other encoding schemes.  We note that the intentionally incomplete population inversion inherent to this scheme does not constitute an error unlike optical losses but is part of the quantum information encoding. More generally, the impact of losses needs to be investigated but does not constitute a systematic limitation. For instance,  photon losses in the present experimental protocol do not impact the fidelity of the spin information measured on the photonic qubit.  As such, we hope that the present work could stimulate the search for new quantum information processing schemes, building on the assets of phase encoding while deriving practical implementations that circumvent the possible impact of photon loss.

\vspace{1em}

\begin{acknowledgments}
EB acknowledges Nune Badalyan for her help in QD spin characterization. This work was conducted within the research program of the QDLight joint laboratory between C2N (UMR 9001 CNRS - Université Paris Saclay - Université Paris Cité) and Quandela SAS. It was partially supported by the Paris Ile-de-France Région in the framework of DIM QUANTIP, the European Union’s Horizon CL4 program under the grant agreement 101135288 for EPIQUE project, ENS Paris-Saclay, the Plan France 2030 through the projects ANR22-PETQ-0011, ANR-22-PETQ-0006, and ANR-22-PETQ-0013, and a public grant overseen by the French National Research Agency as part of the ”Investissements d’Avenir” programme (Labex NanoSaclay, reference: ANR10LABX0035). This work was done within the C2N micro nanotechnologies platforms and partly supported by the RENATECH network and the General Council of Essonne.
\end{acknowledgments}

\bibliography{biblio}

\clearpage
\onecolumngrid
\setcounter{secnumdepth}{3} 
\setcounter{section}{0}
\setcounter{equation}{0}
\setcounter{figure}{0}
\setcounter{table}{0}
\setcounter{page}{1}
\makeatletter
\renewcommand{\thetable}{S\arabic{table}}
\renewcommand{\thesection}{S-\Roman{section}}
\renewcommand{\theequation}{S\arabic{equation}}
\renewcommand{\thefigure}{S\arabic{figure}}
\section*{SUPPEMENTARY INFORMATION AND DATA}

\section{Vacuum--one-photon interference visibility model \label{app_visibility}}
\paragraph{Vacuum--one-photon interference visibility} In this study, interference between consecutive single-rail qubits emitted with a delay $T$, implemented in a path-unbalanced Mach--Zehnder interferometer with a matching internal delay $T$, is used to probe QD spin dynamics encoded in the relative phase between consecutive qubits. We consider each of the Mach-Zehnder interferometer outputs equipped with single-photon detectors (labeled $D_A$ and $D_B$), with respective efficiencies $\eta_A$ and $\eta_B$, measuring single-photon counts $N_x=\eta_x \langle\hat{a}_x^\dagger(t)\hat{a}_x(t)\rangle$.

The associated operators can be expressed in the Heisenberg picture in terms of the input mode $\hat{a}$ by back-propagating the detector modes through the interferometer. 

We first consider the second balanced beamsplitter of the Mach-Zehnder interferometer, which relates the detector modes $\hat{a}_A$ and $\hat{a}_B$ to the modes $\hat{a}_s$ and $\hat{a}_l$ propagating along the short and long Mach-Zehnder interferometer arms, respectively:
\begin{equation}
\hat{a}_A(t) \rightarrow \frac{1}{\sqrt{2}}\left(\hat{a}_s(t)+\hat{a}_l(t)\right), \hspace{1cm}
\hat{a}_B(t) \rightarrow \frac{1}{\sqrt{2}}\left(\hat{a}_s(t)-\hat{a}_l(t)\right).
\end{equation}
We denote by $T$ the delay accumulated in the long arm, matched to the delay between consecutive excitations of the source, and by $\phi$ the internal phase of the Mach-Zehnder interferometer. Attaching this phase to the mode propagating along the long arm, and assuming a balanced first beamsplitter, gives
\begin{equation}
\hat{a}_s(t) \rightarrow \frac{1}{\sqrt{2}}\hat{a}(t), \hspace{1cm}
\hat{a}_l(t) \rightarrow \frac{1}{\sqrt{2}}\hat{a}(t-T)\ep^{\j\phi}.
\end{equation}
Defining $\hat{n}_A(t)=\hat{a}_A^\dagger(t)\hat{a}_A(t)$ and $\hat{n}_B(t)=\hat{a}_B^\dagger(t)\hat{a}_B(t)$, it follows that
\begin{equation}
\hat{n}_A(t)=\frac{1}{4}\left\{\hat{a}^\dagger(t)\hat{a}(t)+\hat{a}^\dagger(t-T)\hat{a}(t-T)+\ep^{\j\phi}\hat{a}^\dagger(t)\hat{a}(t-T)+\ep^{-\j\phi}\hat{a}^\dagger(t-T)\hat{a}(t)\right\},
\end{equation}
\begin{equation}
\hat{n}_B(t)=\frac{1}{4}\left\{\hat{a}^\dagger(t)\hat{a}(t)+\hat{a}^\dagger(t-T)\hat{a}(t-T)-\ep^{\j\phi}\hat{a}^\dagger(t)\hat{a}(t-T)-\ep^{-\j\phi}\hat{a}^\dagger(t-T)\hat{a}(t)\right\}.
\end{equation}

We denote by $\hat{\rho}_{0,T}$ the joint quantum state of the two consecutively emitted photonic wavepackets and by $\hat{\rho}_0$ and $\hat{\rho}_T$ its reduced states at emission times $0$ and $T$. The detector counts are then
\begin{align}
N_A={}&\frac{\eta_A}{4}\int\d t\,\Big[\mathrm{Tr}\!\left(\hat{a}^\dagger(t)\hat{a}(t)\hat{\rho}_T\right)+\mathrm{Tr}\!\left(\hat{a}^\dagger(t-T)\hat{a}(t-T)\hat{\rho}_0\right)+2\,\mathrm{Re}\!\left\{\ep^{\j\phi}\mathrm{Tr}\!\left[\hat{a}^\dagger(t)\hat{a}(t-T)\hat{\rho}_{0,T}\right]\right\}\Big],
\end{align}
\begin{align}
N_B={}&\frac{\eta_B}{4}\int\d t\,\Big[\mathrm{Tr}\!\left(\hat{a}^\dagger(t)\hat{a}(t)\hat{\rho}_T\right)+\mathrm{Tr}\!\left(\hat{a}^\dagger(t-T)\hat{a}(t-T)\hat{\rho}_0\right)-2\,\mathrm{Re}\!\left\{\ep^{\j\phi}\mathrm{Tr}\!\left[\hat{a}^\dagger(t)\hat{a}(t-T)\hat{\rho}_{0,T}\right]\right\}\Big].
\end{align}
Assuming identical photon statistics for the two consecutive emissions, $\hat{\rho}_0=\hat{\rho}_T=\hat{\rho}$, we define
\begin{equation}
I=\int\d t\,\mathrm{Tr}\!\left[\hat{a}^\dagger(t)\hat{a}(t)\hat{\rho}\right],\qquad G^{(1)}(0,T)=\int\d t\,\mathrm{Tr}\!\left[\hat{a}^\dagger(t)\hat{a}(t-T)\hat{\rho}_{0,T}\right].
\end{equation}
The averaged detector counts can then be written
\begin{equation}
N_A=\frac{\eta_A}{2}\left[I+\mathrm{Re}\!\left(\ep^{\j\phi}G^{(1)}(0, T)\right)\right],\hspace{1cm}N_B=\frac{\eta_B}{2}\left[I-\mathrm{Re}\!\left(\ep^{\j\phi}G^{(1)}(0, T)\right)\right]. \label{eqnanb}
\end{equation}
The first term corresponds to the mean emitted photon number, while the second term contains the first-order coherence between the two temporal modes separated by $T$. For equal detector efficiencies,
\begin{equation}
\frac{N_A(\phi)-N_B(\phi)}{N_A(\phi)+N_B(\phi)}=\frac{\mathrm{Re}\!\left[\ep^{\j\phi}G^{(1)}(0, T)\right]}{I}=\mathrm{Re}\left[\ep^{\j\phi} g^{(1)}(0,T)\right],
\end{equation}
where $g^{(1)}(0,T) = G^{(1)}(0, T)/I$.
The interferometer phase and the phase of the photonic coherence therefore enter through their sum. Scanning $\phi$ gives access to the modulus of the coherence, while a phase-stabilized interferometer with a calibrated reference phase could in principle access its phase. In our experiment, the interferometer phase is scanned, and we measure the absolute visibility
\begin{equation}
V=\underset{\phi}{\mathrm{max}}\frac{\left|N_A(\phi)-N_B(\phi)\right|}{N_A(\phi)+N_B(\phi)}=g^{(1)}(0, T).
\end{equation}
For perfectly mode-matched wavepackets restricted to the vacuum--one-photon subspace, $I=\bra{1}\hat{\rho}\ket{1}$ and $G^{(1)}(0, T)=\bra{1_0,0_T}\hat{\rho}_{0,T}\ket{0_0,1_T}$, such that
\begin{equation}
V=\frac{\left|\bra{1_0,0_T}\hat{\rho}_{0,T}\ket{0_0,1_T}\right|}{\bra{1}\hat{\rho}\ket{1}}. \label{visibilityMach-Zehnder interferometer}
\end{equation}

For single-rail qubits of single-photon population $p=\bra{1}\hat{\rho}\ket{1}$, we define the intrinsic vacuum--one-photon interference visibility $V_0$ by $V\equiv V_0(1-p)$, such that
\begin{equation}
V_0=\frac{\left|\bra{1_0,0_T}\hat{\rho}_{0,T}\ket{0_0,1_T}\right|}{p(1-p)}. \label{visibilityV0}
\end{equation}
This normalization removes the visibility reduction associated with the finite single-photon preparation probability $p$, such that $V_0=1$ for two identical single-rail qubits interfering.

\paragraph{Photon loss}

Importantly, this visibility measurement is insensitive to identical linear photon losses acting on the two temporal modes after state generation. Considering a loss channel of transmission $x$, the one-photon population is reduced as $\bra{1}\hat{\rho}\ket{1}\rightarrow x\bra{1}\hat{\rho}\ket{1}$, while the two-time coherence entering Eq.~\ref{visibilityMach-Zehnder interferometer} transforms as $\bra{1_0,0_T}\hat{\rho}_{0,T}\ket{0_0,1_T}\rightarrow x\bra{1_0,0_T}\hat{\rho}_{0,T}\ket{0_0,1_T}$. Consequently,
\begin{equation}
V_x=\frac{x\left|\bra{1_0,0_T}\hat{\rho}_{0,T}\ket{0_0,1_T}\right|}{x\bra{1}\hat{\rho}\ket{1}}=V.
\end{equation}
vacuum--one-photon interference visibility therefore naturally normalizes out identical linear losses. In the definition $V=V_0(1-p)$, $p$ consequently refers to the single-photon preparation probability at the source, before such propagation or collection losses.

\paragraph{Effect of partial photon distinguishability}
Partial single-photon distinguishability reduces the overlap between the two interfering temporal modes. Denoting the single-photon indistinguishability by $M_s$, the field-coherence amplitude entering Eq.~\ref{eqnanb} is reduced by $\sqrt{M_s}$~\cite{Loredo_Antón_Reznychenko_Hilaire_Harouri_Millet_Ollivier_Somaschi_DeSantis_Lemaître_etal._2019}. Together with the finite single-photon population $p$, the experimentally measured visibility is therefore
\begin{equation}
V=V_0(1-p)\sqrt{M_s}. \label{visibility_distinguishability}
\end{equation}
The independent characterization of $p$ and $M_s$ described above therefore allows the intrinsic visibility $V_0$ to be extracted from the measured interference contrast.

\paragraph{Single-photon purity} 
As shown in \cite{Wenniger_Wein_Fioretto_Thomas_Antón-Solanas_Lemaître_Sagnes_Harouri_Belabas_Somaschi_etal._2024}, partial single-photon purity affects the first-order coherence between the two temporal modes of the pure separate single-rail qubits according to $g^{(1)}(0, T) = (1-p) + 2p\sqrt{(1-p)g^{(2)}(0)}$, assuming $g^{(2)}(0) \ll 1$ limit. As a consequence, the corrections are only minor, and the leading term in the visibility dependence on $p$ is given by the pure single photons, as used in the main text.


\section{Vacuum--one-photon interference measurement of separable and correlated states}

\paragraph{Separate photonic states} In the absence of correlations between consecutive emissions, $\hat{\rho}_{0,T}=\hat{\rho}_0\otimes\hat{\rho}_T$. For identical stationary states $\hat{\rho}_0=\hat{\rho}_T=\hat{\rho}$, Eq.~\ref{visibilityV0} reduces to
\begin{equation}
V_0=\frac{\left|\bra{0}\hat{\rho}\ket{1}\right|^2}{p(1-p)}.
\end{equation}
In this case, $V_0$ no longer probes correlations between two emissions but directly measures the normalized photon-number coherence of the individual emitted qubits $C_0$, defined based on the first-order photon-number coherence $c^{(1)} \equiv \frac{\left|\bra{0}\hat{\rho}\ket{1}\right|^2}{p} \equiv C_0(1-p)$. With these notations, a general state can be conveniently written
\begin{equation}
\hat{\rho}=\begin{pmatrix}1-p & \sqrt{C_{\mathrm{0}}}\,\ep^{-\j\varphi}\sqrt{p(1-p)}\\ \sqrt{C_{\mathrm{0}}}\,\ep^{\j\varphi}\sqrt{p(1-p)} & p\end{pmatrix}.
\end{equation}

\paragraph{A phase parity measurement} The intrinsic visibility $V_0$ measures the normalized two-time photon-number coherence between the two consecutively emitted photonic qubits. Its relation to phase parity becomes particularly transparent for statistical mixtures of phase-labeled states defined throughout this work as $\ket{\varphi_p^\pm}\equiv\sqrt{1-p}\ket{0}+\ep^{\j(\varphi\pm\frac{\pi}{2})}\sqrt{p}\ket{1}$. The two states have identical photon-number populations but opposite vacuum--one-photon coherences.

First, consider two consecutively emitted phase-labelled states $\ket{\varphi_p^{s_0}}^0$ and $\ket{\varphi_p^{s_T}}^T$, with the phase labels $s_0,s_T=\pm1$. Their contribution to the coherence entering Eq.~\ref{visibilityV0} is
\begin{align}
\bra{1_0,0_T}\left(\ket{\varphi_p^{s_0}}\bra{\varphi_p^{s_0}}\otimes\ket{\varphi_p^{s_T}}\bra{\varphi_p^{s_T}}\right)\ket{0_0,1_T}&=\left[\ep^{\j(\varphi+s_0\frac{\pi}{2})}\sqrt{p}\right]\sqrt{1-p}\sqrt{1-p}\left[\ep^{-\j(\varphi+s_T\frac{\pi}{2})}\sqrt{p}\right]\notag\\
&=s_0s_T\,p(1-p).
\end{align}
The common phase $\varphi$ cancels, and the sign of the coherence is therefore determined by the relative phase parity of the two states, reflected by the product of phase-labels $s_0 s_T$: it is positive for equal parities $(++$ or $--)$ and negative for opposite parities $(+-$ or $-+)$. Consequently, for a statistical mixture of such phase-labelled product states
\begin{equation}
V_0=\left|\left\langle s_0s_T\right\rangle\right|,  \label{eq_phaseparity}
\end{equation}
and the vacuum--one-photon interference visibility directly measures the modulus of the phase-parity correlation.

\section{Self-homodyne visibility of spin-photon phase entanglement}

In the case of single-rail qubit generation by an ideal charged exciton source, consecutive qubits cannot, in general, be treated as separable states.

\paragraph{Initial pure spin state}
We first consider the most general pure QD spin state $\sqrt{p_\ua}\ket{\ua}_z+\ep^{\j\chi}\sqrt{p_\da}\ket{\da}_z$, with $p_\ua+p_\da=1$, where $\chi$ is the relative phase between the two spin components. Under $H$-polarized coherent excitation and upon consecutive spontaneous emission at time labeled $0$, the spin-photonic qubit is given by $\sqrt{p_\ua}\ket{\ua}_z\ket{\varphi_p^-}^0+\ep^{\j\chi}\sqrt{p_\da}\ket{\da}_z\ket{\varphi_p^+}^0$. In our Voigt-geometry configuration, the magnetic field is oriented along $y$, such that, in the absence of spin decoherence, the evolution during the delay $T$ corresponds to a rotation of angle $\theta_T=\omega_L T$:
\begin{equation}
\ket{\ua}_z\rightarrow\cos\frac{\theta_T}{2}\ket{\ua}_z+\sin\frac{\theta_T}{2}\ket{\da}_z,\qquad
\ket{\da}_z\rightarrow-\sin\frac{\theta_T}{2}\ket{\ua}_z+\cos\frac{\theta_T}{2}\ket{\da}_z.
\end{equation}
After spin rotation and second excitation at time $T$, the state describing the spin and the two photonic emitted wavepackets is
\begin{align}
\ket{\psi}={}&\ket{\ua}_z\ket{\varphi_p^-}^T
\left[
\sqrt{p_\ua}\cos\frac{\theta_T}{2}\ket{\varphi_p^-}^0
-\sqrt{p_\da}\ep^{\j\chi}\sin\frac{\theta_T}{2}\ket{\varphi_p^+}^0
\right]\notag\\
&+\ket{\da}_z\ket{\varphi_p^+}^T
\left[
\sqrt{p_\ua}\sin\frac{\theta_T}{2}\ket{\varphi_p^-}^0
+\sqrt{p_\da}\ep^{\j\chi}\cos\frac{\theta_T}{2}\ket{\varphi_p^+}^0
\right].
\end{align}
Tracing over the final spin state gives
\begin{align}
\hat{\rho}_{0,T}={}&
\ket{\varphi_p^-}^T\bra{\varphi_p^-}^T
\Bigg[
p_\ua\cos^2\frac{\theta_T}{2}\ket{\varphi_p^-}^0\bra{\varphi_p^-}^0
+p_\da\sin^2\frac{\theta_T}{2}\ket{\varphi_p^+}^0\bra{\varphi_p^+}^0
\Bigg]\notag\\
&+\ket{\varphi_p^+}^T\bra{\varphi_p^+}^T
\Bigg[
p_\ua\sin^2\frac{\theta_T}{2}\ket{\varphi_p^-}^0\bra{\varphi_p^-}^0
+p_\da\cos^2\frac{\theta_T}{2}\ket{\varphi_p^+}^0\bra{\varphi_p^+}^0
\Bigg]\notag\\
&+\sqrt{p_\ua p_\da}\,\frac{\sin\theta_T}{2}
\left(
-\ket{\varphi_p^-}^T\bra{\varphi_p^-}^T
+\ket{\varphi_p^+}^T\bra{\varphi_p^+}^T
\right) \otimes
\left[
\ep^{-\j\chi}\ket{\varphi_p^-}^0\bra{\varphi_p^+}^0
+\ep^{\j\chi}\ket{\varphi_p^+}^0\bra{\varphi_p^-}^0
\right].
\end{align}
Using Eq.~\ref{eq_phaseparity}, the intrinsic vacuum--one-photon interference visibility is therefore
\begin{equation}
V_0
=
\sqrt{
\cos^2\theta_T
+
4p_\ua p_\da\sin^2\theta_T\sin^2\chi
}.
\label{eq_v0purenodec}
\end{equation}
For $\chi=0$ or $\pi$, the second term vanishes and $V_0=|\cos\theta_T|$. Conversely, for a spin initially prepared along $y$, $p_\ua=p_\da=1/2$ and $\chi=\pm\pi/2$, yielding $V_0=1$ independently of $\theta_T$: the spin rotation then only changes the phase of the interference fringes.

\paragraph{Initial mixed spin state}
In our experiment, the spin is not initialized in a reproducible pure state. We therefore consider an incoherent initial spin mixture in the $\{\ket{\ua}_z,\ket{\da}_z\}$ basis,
\begin{equation}
\hat{\rho}_s
=
p_\ua\ket{\ua}_z\bra{\ua}_z
+
p_\da\ket{\da}_z\bra{\da}_z,
\end{equation}
with $p_\ua+p_\da=1$. In the absence of residual spin polarization, one expects $p_\ua=p_\da=1/2$, corresponding to the completely unpolarized state $\hat{\rho}_s=\mathbb{I}/2$. We nevertheless keep $p_\ua$ and $p_\da$ general to account for a possible residual population imbalance. After the first emission,
\begin{equation}
\hat{\rho}_{s,0}
=
p_\ua\ket{\ua}_z\bra{\ua}_z\otimes
\ket{\varphi_p^-}^0\bra{\varphi_p^-}^0
+
p_\da\ket{\da}_z\bra{\da}_z\otimes
\ket{\varphi_p^+}^0\bra{\varphi_p^+}^0.
\end{equation}
The two components of this statistical mixture can be propagated independently. After spin rotation and second excitation at time $T$ followed by photon emission,
\begin{align}
\ket{\Psi_\ua}
&=
\cos\frac{\theta_T}{2}
\ket{\ua}_z\ket{\varphi_p^-}^T\ket{\varphi_p^-}^0
+
\sin\frac{\theta_T}{2}
\ket{\da}_z\ket{\varphi_p^+}^T\ket{\varphi_p^-}^0, \notag \\
\ket{\Psi_\da}
&=
-\sin\frac{\theta_T}{2}
\ket{\ua}_z\ket{\varphi_p^-}^T\ket{\varphi_p^+}^0
+
\cos\frac{\theta_T}{2}
\ket{\da}_z\ket{\varphi_p^+}^T\ket{\varphi_p^+}^0.
\end{align}
The complete state after the two excitations is therefore
\begin{equation}
\hat{\rho}_{s,0,T}
=
p_\ua\ket{\Psi_\ua}\bra{\Psi_\ua}
+
p_\da\ket{\Psi_\da}\bra{\Psi_\da}.
\end{equation}
Tracing over the final spin state gives
\begin{align}
\hat{\rho}_{0,T}={}&
\ket{\varphi_p^-}^T\bra{\varphi_p^-}^T
\left[
p_\ua\cos^2\frac{\theta_T}{2}
\ket{\varphi_p^-}^0\bra{\varphi_p^-}^0
+
p_\da\sin^2\frac{\theta_T}{2}
\ket{\varphi_p^+}^0\bra{\varphi_p^+}^0
\right]\notag\\
&+
\ket{\varphi_p^+}^T\bra{\varphi_p^+}^T
\left[
p_\ua\sin^2\frac{\theta_T}{2}
\ket{\varphi_p^-}^0\bra{\varphi_p^-}^0
+
p_\da\cos^2\frac{\theta_T}{2}
\ket{\varphi_p^+}^0\bra{\varphi_p^+}^0
\right].
\end{align}
Unlike for an initial coherent spin superposition, no terms coupling $\ket{\varphi_p^-}$ and $\ket{\varphi_p^+}$ appear, since there is initially no coherence between $\ket{\ua}_z$ and $\ket{\da}_z$. Using Eq.~\ref{eq_phaseparity},
\begin{equation}
V_0
=
\left|
p_\ua\cos^2\frac{\theta_T}{2}
-p_\da\sin^2\frac{\theta_T}{2}
-p_\ua\sin^2\frac{\theta_T}{2}
+p_\da\cos^2\frac{\theta_T}{2}
\right|=
|\cos\theta_T|.
\end{equation}
The result is therefore independent of the initial spin populations. In particular, the completely unpolarized case $p_\ua=p_\da=1/2$ results in the identical expression. This coincides with Eq.~\ref{eq_v0purenodec} when the initial coherent contribution vanishes.

\paragraph{Visibility and spin correlation}
More in general, for an initial mixed spin state, $V_0$ measures the temporal correlation of the spin along $z$. The spin-photon mapping $\ket{\uparrow(\downarrow)}_{z}\bra{\uparrow(\downarrow)}_{z}\rightarrow \ket{\varphi_{p}^{-(+)}}\bra{\varphi_{p}^{-(+)}}$  directly relates the photonic parity $s$ to the eigenvalue $S_z$ of operator $\hat{S}_z$ projecting the spin state along $z$ direction: $\ket{\ua}_z$, with $S_z=+1$, generates $\ket{\varphi_p^-}$, corresponding to $s=-1$; while $\ket{\da}_z$, with $S_z=-1$, generates $\ket{\varphi_p^+}$, corresponding to $s=+1$. Therefore, at each emission time, $s=-S_z$, and consequently $s_0s_T=\left[-S_z(0)\right]\left[-S_z(T)\right]=S_z(0)S_z(T)$. Plugging this replacement in Eq.~\eqref{eq_phaseparity} therefore gives
\begin{equation}
V_0(T)= \left|\left\langle\hat{S}_z(T)\hat{S}_z(0)\right\rangle\right| \label{eq_v0mixeddec}
\end{equation}
Let us notice that the above expression is valid also in the case where the spin undergoes decoherence, in this case a spin initialized in the pure state $\ket{\uparrow(\downarrow)}$ evolves into the mixed spin state $\rho_{s}^{\uparrow(\downarrow)}(T)$. Equation~\eqref{eq_v0mixeddec} can be rewritten as $V_0=\left| p_{\uparrow}\text{Tr}[\hat{S}_{z},\rho^{\uparrow}(T)]-p_{\downarrow}\text{Tr}[\hat{S}_{z},\rho^{\downarrow}(T)]\right|$ and the average value $\text{Tr}[\hat{S}_{z},\rho^{\uparrow}(T)]=-\text{Tr}[\hat{S}_{z},\rho^{\downarrow}(T)]$ can be obtained from the Merkulov-Efros-Rosen spin-decoherence model~\cite{Merkulov_Efros_Rosen_2002} following:
\begin{equation}
\begin{aligned}
\left\langle\hat{S}_z(T)\hat{S}_z(0)\right\rangle =\;&
\exp\left[-\left(\frac{t}{2T_O}\right)^2\right]
\left[
\cos\left(\frac{Bt}{\Delta_B T_O}\right)
-\frac{\Delta_B t}{2BT_O}
\sin\left(\frac{Bt}{\Delta_B T_O}\right)
\right]
\\
&+\frac{1}{2\sqrt{\pi}}
\int_{-\infty}^{+\infty}\mathrm{d}z
\int_0^{+\infty}\mathrm{d}y\,
\frac{y\exp(-z^2-y)}
{\left(\frac{B}{\Delta_B}+z\right)^2+y} \times
\left[
1-\cos\left(
\sqrt{\left(\frac{B}{\Delta_B}+z\right)^2+y}
\frac{t}{T_O}
\right)
\right].
\end{aligned}
\end{equation}

\section{concurrence of spin-photon phase entanglement \label{app_concurrence}}

The phase-labelled states $\ket{\varphi_p^-}$ and $\ket{\varphi_p^+}$ correspond to phase-anticorrelated states with opposite relative phases, with vacuum--one-photon coherences differing by a phase $\pi$. Their overlap is $\braket{\varphi_p^-|\varphi_p^+}=1-2p$, such that they are orthogonal for the balanced single-rail qubit $p=1/2$. Away from this balanced case, their partial overlap reduces the achievable spin--photon concurrence by a factor $\sqrt{1-\left|\braket{\varphi_p^-|\varphi_p^+}\right|^2}=2\sqrt{p(1-p)}$. In the following, we therefore assume $p=1/2$.

\paragraph{Initial $\ket{\ua}_z$ spin}
For a QD spin initialized in $\ket{\ua}_z$, an arbitrary spin rotation of angle $\theta_T$ around the $y$ axis prepares the state $\cos(\frac{\theta_T}{2})\ket{\ua}_z+\sin(\frac{\theta_T}{2})\ket{\da}_z$. Coherent excitation then encodes the two spin components onto the two states with oposite phase of the emitted single-rail qubit, yielding $\cos\frac{\theta_T}{2}\ket{\ua}_z\ket{\varphi_{1/2}^-}+\sin\frac{\theta_T}{2}\ket{\da}_z\ket{\varphi_{1/2}^+}$. In the joint basis  $\left\{\ket{\ua}_z\ket{\varphi_{1/2}^-},\ket{\ua}_z\ket{\varphi_{1/2}^+},\ket{\da}_z\ket{\varphi_{1/2}^-},\ket{\da}_z\ket{\varphi_{1/2}^+}\right\}$, the corresponding density matrix writes
\begin{equation}
\hat{\rho}_T=
\begin{pmatrix}
\cos^2\frac{\theta_T}{2} & 0 & 0 & \frac{\sin\theta_T}{2}\\
0 & 0 & 0 & 0\\
0 & 0 & 0 & 0\\
\frac{\sin\theta_T}{2} & 0 & 0 & \sin^2\frac{\theta_T}{2}
\end{pmatrix}.
\end{equation}
The corresponding spin-photon concurrence is $\mathcal{C}(T)=|\sin\theta_T|$, maximal for a $\pi/2$ spin rotation $\mathcal{C}(T_{\pi/2})=1$.

\paragraph{Initial mixed spin} 
In the present experiment, the QD spin is not initialized before the spin rotation. In the absence of residual spin polarization, the initial spin state is the maximally mixed state $\hat{\rho}_s=\mathbb{I}/2$. In the basis defined above, this state writes
\begin{equation}
\hat{\rho}_T=\frac{1}{2}
\begin{pmatrix}
1 & 0 & 0 & 0\\
0 & 0 & 0 & 0\\
0 & 0 & 0 & 0\\
0 & 0 & 0 & 1
\end{pmatrix},
\end{equation}
showing no spin-photon phase entanglement, $\mathcal{C}(T)=0$. Spin-photon phase entanglement is therefore discussed in this work for an initialized QD spin.

\paragraph{Decohering initialized spin} 
We finally consider a spin undergoing decoherence during a time $T$ before coherent excitation, a spin initialized in the pure state $\ket{\uparrow(\downarrow)}_{z}$ evolves into the mixed spin state $\rho_{s}^{\uparrow(\downarrow)}(T)$. The average values of the spin components $\langle \hat{S}_{j}(T)\rangle_{\uparrow(\downarrow)}=\text{Tr}[\hat{S}_{j},\rho_{s}^{\uparrow(\downarrow)}(T)]$ with $j=x,y,z$, can be computed from the Merkulov spin-decoherence model~\cite{Merkulov_Efros_Rosen_2002}. Therefore, if the spin initial state is $\ket{\uparrow}_{z}$, the single-rail qubit emission maps this state onto the spin-photon density matrix
\begin{equation}
\hat{\rho}_T=\frac{1}{2}\begin{pmatrix}1+\langle\hat{S}_z(T)\rangle_{\uparrow} & 0 & 0 & \langle\hat{S}_x(T)\rangle_{\uparrow}\\ 0 & 0 & 0 & 0\\ 0 & 0 & 0 & 0\\ \langle\hat{S}_x(T)\rangle_{\uparrow} & 0 & 0 & 1-\langle\hat{S}_z(T)\rangle_{\uparrow}\end{pmatrix},
\end{equation}
written in the basis defined above. Its concurrence is $\mathcal{C}(T)=\left| \langle\hat{S}_x(T)\rangle_{\uparrow}\right|=\left| \langle\hat{S}_x(T)\rangle_{\downarrow}\right|$ which follows the decoherence model~\cite{Merkulov_Efros_Rosen_2002}:
\begin{equation}
\left|\langle\hat{S}_x(T)\rangle_{\uparrow}\right|
= \frac{1}{\sqrt{\pi}}
\int_{-\infty}^{+\infty}\!\mathrm{d}z
\int_0^{+\infty}\!\mathrm{d}y\,
\frac{\left[(B/\Delta_B+z)^3+zy\right]\mathrm{e}^{-z^2-y}}
{\left[(B/\Delta_B+z)^2+y\right]^{3/2}}
\sin\left(\sqrt{(B/\Delta_B+z)^2+y}\,\frac{T}{T_O}\right).
\end{equation}

\section{Accessing spin-photon phase entanglement concurrence}

Although not implemented, we assume the QD was initially prepared in $\ket{\ua}_z$. As previously shown, a spin rotation of angle $\theta_T$ followed by coherent excitation generates entanglement between the semiconductor spin and the phase of the emitted single-rail qubit. For balanced qubits with $p=1/2$ and in our geometry, where $\langle\hat{S}_y(T)\rangle=0$, the spin-photon phase entanglement concurrence is directly given by $\mathcal{C}(T)=|\langle\hat{S}_x(T)\rangle|$, while the vacuum--one-photon interference visibility is $V_0(T)=|\langle\hat{S}_z(T)\rangle|$. For an ideal coherent spin evolution, the Bloch-vector length is preserved, such that
\begin{equation}
\mathcal{C}^2(T)+V_0^2(T)=\langle\hat{S}_x(T)\rangle^2+\langle\hat{S}_z(T)\rangle^2=1.
\end{equation}

\begin{figure}[H]
\centering
\includegraphics[scale=0.58]{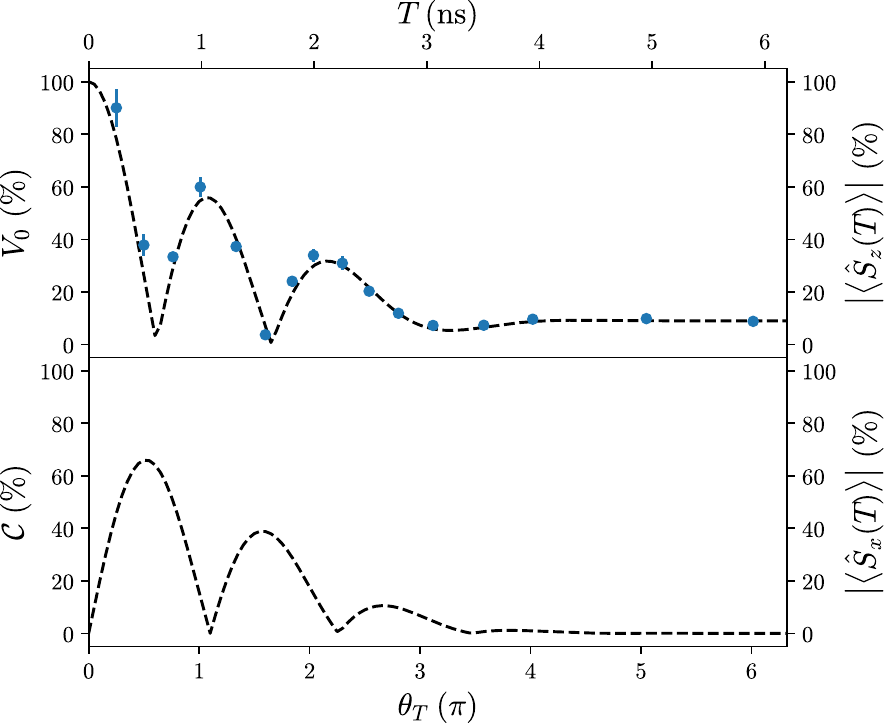}
\caption{\textbf{Towards spin-photon phase entanglement generation.} Spin-photon concurrence $\mathcal{C}(T)$ expected for a spin initially prepared in $\ket{\ua}_z$ and a balanced single-rail qubit with $p=1/2$ is reconstructed in the lower panel from vacuum--one-photon interference visibility measurement at $B=60~\rm mT$ fitted using Merkulov spin decoherence model~\cite{Merkulov_Efros_Rosen_2002}. }
\label{fig_merkulovconcurrence}
\end{figure}

In a semiconductor QD, however, the central spin is subject to decoherence, characterized here by the Overhauser-induced dephasing timescale $T_{\mathrm O}=0.88\pm0.01~\mathrm{ns}$. The resulting reduction of the Bloch-vector length modifies the previous relation to
\begin{equation}
\mathcal{C}^2(T)+V_0^2(T)=\langle\hat{S}_x(T)\rangle^2+\langle\hat{S}_z(T)\rangle^2<1.
\end{equation}
The measured vacuum--one-photon interference visibility therefore provides $|\langle\hat{S}_z(T)\rangle|$ and allows extraction of the parameters of the Merkulov model. The same model gives the predicted entanglement concurrence $\mathcal{C}(T)=|\langle\hat{S}_x(T)\rangle|$, shown in Fig.~\ref{fig_merkulovconcurrence}, illustrating periodic transfer of quantum information between visibility and concurrence, dissipated by the QD spin. In particular, assuming no QD spin decoherence, maximal concurrence is expected for $\theta_T=\frac{\pi}{2}$ up to $\pi$, corresponding to minimal visibilities. Assuming additional spin initialization, a maximum predicted concurrence of $\mathcal{C}=65.9\%$ is obtained for $T=0.49\rm ns$ and $B=60\rm mT$.

\section{Experimental measurement \label{app_expvisibility}}

\paragraph{QD-based source}
All measurements were performed using a semiconductor InGaAs QD-based single-photon source produced by Quandela. 
The QD is grown by molecular beam epitaxy in the Stranski-Krastanov regime and is embedded between two distributed Bragg reflectors manufactured by alternating GaAs/AlGaAs pairs, forming an asymmetric cavity with 18 pairs on top (36 pairs below) the undoped GaAs $\lambda$-cavity region, such that the QD is in the node of its optical field. The top (bottom) mirrors are n (p) doped, forming a p-i-n diode, to provide electrical control of the QD charge state and its resonance energy. Additionally, we laterally confine the light by etching a cylindrical (slightly birefringent, splitting $31.4~\rm\mu$s) micropillar of a few microns in diameter and a quality factor $Q=\left(10.0\pm0.1\right)\rm k$, deterministically centered around the QD. The QD, cooled to 4 K, is tuned electrically to its singly-charged state and to resonance with the fundamental micropillar cavity mode around 925\,nm. 

\begin{figure}[H]
\centering
\includegraphics[width=\textwidth]{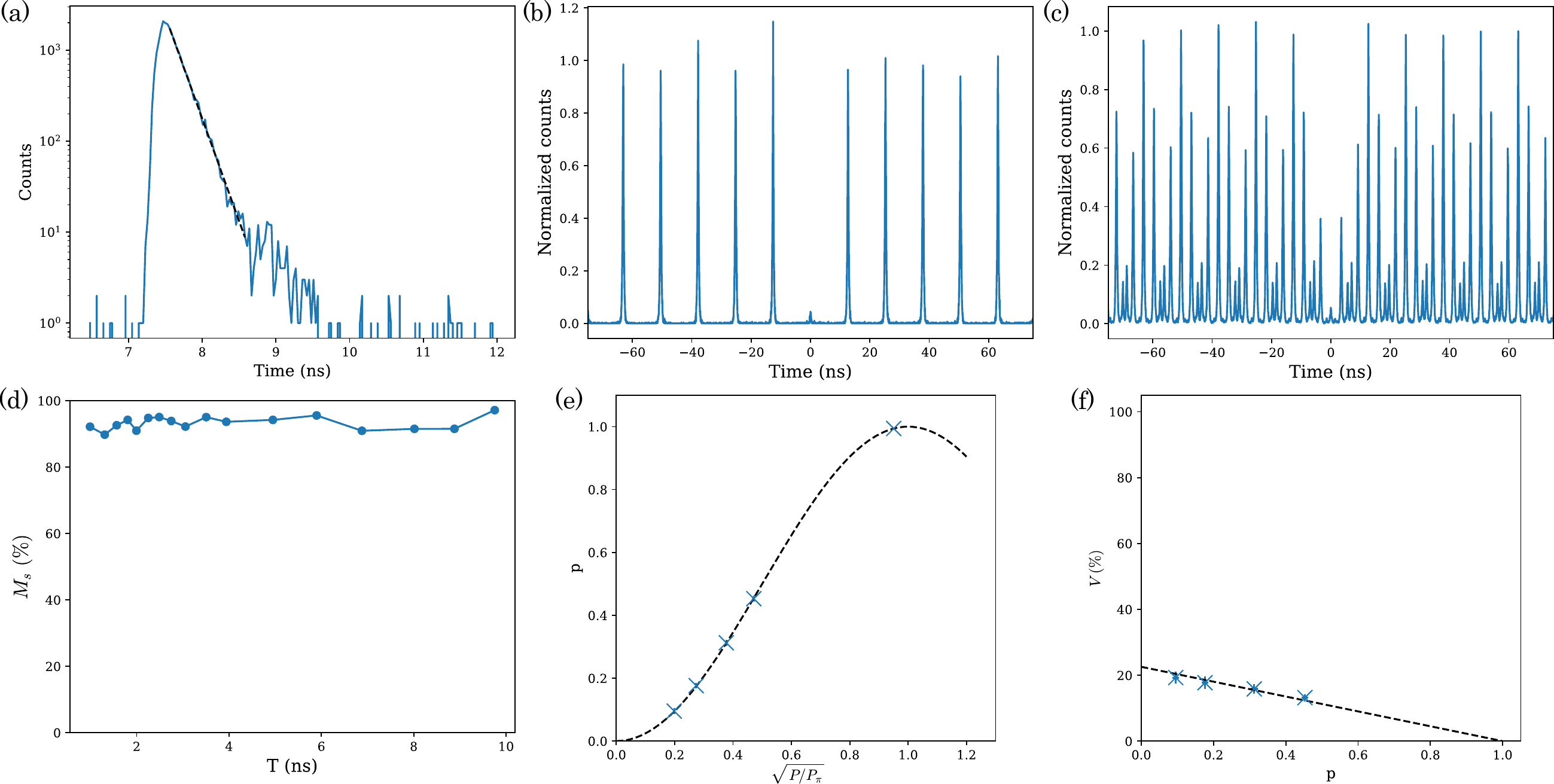}
\caption{\textbf{Single-photon and single-rail qubit characterization.} Single-photon source properties are shown in panels \textbf{(a--c)}, displaying the emission lifetime of the negatively charged exciton transition, single-photon purity, and HOM measurement, respectively, with $\tau=193\pm4~\rm ps$, $g^{(2)}(0)=4.13\pm0.06\%$ and $M_s=95.04\pm0.19\%$. Photon indistinguishability measurement was repeated for various delays $T>1~\rm ns$, yielding $M_s=93.2\pm0.2~\%$ on average. The single-rail qubit population $p$ is calibrated from Rabi oscillations of the QD signal intensity as a function of the normalized drive amplitude $\sqrt{P/P_\pi}$ in \textbf{(e)}. An example of a vacuum--one-photon interference visibility measurement is shown in \textbf{(f)}.}
\label{fig_perfsource}
\end{figure}

\begin{figure}[H]
\centering
\includegraphics[width=0.6\textwidth]{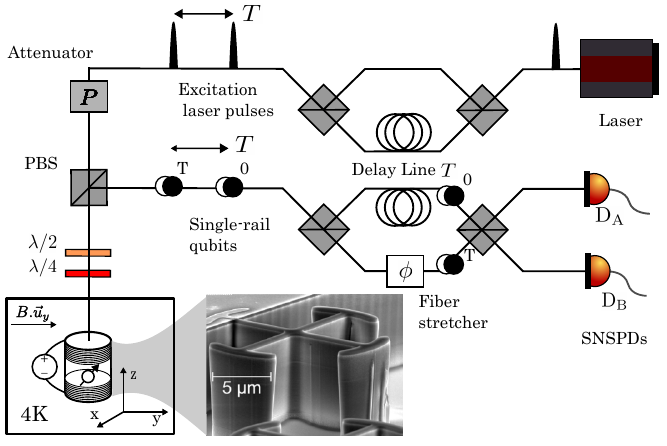}
\caption{\textbf{Pump-probe generation and vacuum--one-photon interference measurement.} Each excitation pulse is split into two pulses separated by a delay $T$ using a first Mach--Zehnder interferometer, sequentially generating two single-rail qubits. vacuum--one-photon interference is subsequently measured using a second Mach--Zehnder interferometer with the same internal delay $T$. The excitation sequence is repeated every $12.3~\mathrm{ns}$.}
\label{fig_setupcomplete}
\end{figure}

The device is operated with $\approx10$ ps long resonant linearly-polarized Gaussian pulses, which are separated from the reflected excitation laser using a cross-polarization technique in a confocal microscope~\cite{Steindl_Frey_Norman_Bowers_Bouwmeester_Löffler_2023}. The otpical performances of the source relevant to the vacuum--one-photon interference measurements are summarized in Fig.~\ref{fig_perfsource}. In particular, panels \textbf{(a--c)} characterize the optical properties of the emitted photons through the photon lifetime of exponential decay of  $193\pm4$\,ps, the second-order correlation function, and Hong-Ou-Mandel interference, respectively. We measure a high single-photon purity $1-g^{(2)}(0)=95.87\pm0.06\%$ and a single-photon indistinguishability $M_s=95.04\pm0.19\%$ at $T=3.5~\rm ns$. The photon indistinguishability measurement shown in panel \textbf{(c)} exhibits repeated groups of five peaks, arising from the correlation measurement performed over three temporal modes using the experimental apparatus shown in Fig.~\ref{fig_setupcomplete}. This measurement was repeated for delays $T>1~\rm ns$, as shown in panel \textbf{(d)}, yielding consistently high photon indistinguishability with an average value $M_s=93.2\pm0.2\%$.

The coherent excitation of the charged exciton also allows the emitted state to be continuously tuned between vacuum and one photon. The corresponding single-photon population $p=\sin^2\sqrt{P/P_\pi}$ is calibrated from Rabi oscillations of the QD intensity as a function of the normalized drive amplitude $\sqrt{P/P_\pi}$, shown in Fig.~\ref{fig_perfsource}\textbf{(e)}. This calibration is required to extract the intrinsic visibility $V_0$ from the measured vacuum--one-photon interference visibility, since the finite vacuum component reduces the measured visibility by a factor $1-p$. The independently measured indistinguishability $M_s$ provides the second correction entering the experimental extraction of $V_0$. Figure~\ref{fig_perfsource}\textbf{(f)} shows a representative vacuum--one-photon interference visibility measurement.

\paragraph{Pump-probe vacuum--one-photon interference measurement}

The vacuum--one-photon interference measurement introduced in Ref.~\cite{Loredo_Antón_Reznychenko_Hilaire_Harouri_Millet_Ollivier_Somaschi_DeSantis_Lemaître_etal._2019} is implemented here in a pump-probe configuration. A first Mach--Zehnder interferometer splits each excitation pulse into two pulses separated by a controllable delay $T$. These pulses sequentially excite the QD and generate two single-rail qubits separated by the same delay. The emitted qubits are subsequently sent to a second Mach--Zehnder interferometer whose internal delay is matched to $T$, so that the long-path contribution of the first emission temporally overlaps with the short-path contribution of the second emission.

The internal phase $\phi$ of the second interferometer is modulated using a fiber stretcher at a frequency close to $0.5~\mathrm{Hz}$. For each delay $T$, the interference visibility $V$ is measured for several populations $p$. The intrinsic visibility $V_0$ is then obtained from Eq.~\ref{visibility_distinguishability}. Varying $T$ therefore directly probes the decay and oscillation of the phase correlation between consecutive emitted qubits and, through the spin--photon information transfer derived above, the underlying two-time QD spin correlation.

\paragraph{Additional interference phenomena}
The idealized two-mode description introduced above assumes that only the two wavepackets that overlap at the final beamsplitter contribute to the interference signal. In the pump-probe implementation, however, the two emitted qubits generate three temporal output windows. The first qubit, emitted at time $t-T$, exits the analysis interferometer at either $t-T$ or $t$, while the second qubit, emitted at time $t$, exits at either $t$ or $t+T$. The corresponding output field operators can be written
\begin{align}
\hat{a}^{\mathrm{PP}}_{A}(t)&\rightarrow\frac{1}{2}\left(\ep^{\j\phi}\hat{a}^{t+T}(t)+\hat{a}^{t}(t)+\ep^{\j\phi}\hat{a}^{t}(t-T)+\hat{a}^{t-T}(t-T)\right), \label{eq_na2}\\
\hat{a}^{\mathrm{PP}}_{B}(t)&\rightarrow\frac{1}{2}\left(-\ep^{\j\phi}\hat{a}^{t+T}(t)+\hat{a}^{t}(t)-\ep^{\j\phi}\hat{a}^{t}(t-T)+\hat{a}^{t-T}(t-T)\right). \label{eq_nb2}
\end{align}
Here, the argument indicates the emission time of the photonic wavepacket, while the superscript labels the corresponding output temporal window. Only the two contributions in the central temporal window, $\hat{a}^{t}(t)$ and $\hat{a}^{t}(t-T)$, carry the vacuum--one-photon interference between consecutive emissions.

When the photon wavepacket duration is much shorter than $T$, the three temporal windows are well separated. The central interfering window then accounts for half of the total detected intensity, while the two outer windows provide a non-interfering background. The experimentally observed pump-probe visibility is consequently
\begin{equation}
V^{\mathrm{PP}}=\frac{V}{2}.
\end{equation}

The delay $T$ is scanned from $250\,\mathrm{ps}$ to $10\,\mathrm{ns}$. At short delays, $T\leq1\,\mathrm{ns}$, the finite temporal extent of the photon wavepackets causes the three output windows to partially overlap. A given wavepacket can then interfere with itself through the short and long interferometer paths, adding a which-path interference contribution $V_{\mathrm{WPI}}$ to the pump-probe signal. This contribution is independently measured using a single excitation pulse, allowing the total measured visibility to be written
\begin{equation}
V^{\mathrm{PP}}=V_{\mathrm{WPI}}+\frac{V}{2},
\end{equation}
from which the vacuum--one-photon interference visibility $V$ can be extracted.

A second effect appears in the same short-delay regime. In addition to HOM, vacuum--one-photon, and which-path interference, the two coherent excitation pulses can drive Ramsey interference of the QD populations~\cite{Wenniger_Maffei_Wein_Prasad_Lam_Fioretto_Lemaître_Sagnes_Antón-Solanas_Senellart_etal._2026}. This contribution is not included in the present model and constitutes the dominant additional uncertainty in the extraction of $V_0$ at the shortest delays.

\section{Additional spin measurements \label{app_spincorrelation}}

The interference between the two single-rail qubits converts the phase correlation between consecutive photonic qubits into the visibility $V_0$. As shown in the previous section, for the non-initialized spin considered experimentally, $V_0$ effectively quantifies the two-time spin correlation between time $t$ and $t+T$
\begin{equation}
V_0(T)=\left|\left\langle\hat{S}_z(T)\hat{S}_z(0)\right\rangle\right|.
\end{equation}
The same spin dynamics can therefore be investigated either by varying the temporal separation $T$ of the probe pulses at fixed magnetic field, as in the measurements presented in the main text, or by varying the magnetic field to control the spin precession at fixed probing times $T$. The latter approach provides a complementary view of the same spin-dephasing mechanisms, as used below.

\paragraph{Hanle curves}
Complementary to the time-dependent Merkulov curves shown in Fig.~\ref{fig_merkulov}, we fix the probing delay $T$ and vary the applied in-plane magnetic field $B$, in a Hanle-type experiment \cite{Bracker_Stinaff_Gammon_Ware_Tischler_Shabaev_Efros_Park_Gershoni_Korenev_etal._2005}, thereby tuning both the Larmor frequency and the relative importance of the external and Overhauser fields. For $T=1.55~\mathrm{ns}$, shown in panel \textbf{(a)}, comparable to the Overhauser-induced dephasing timescale, $V_0$ exhibits oscillations as a function of $B$, reflecting the coherent spin precession while its amplitude is progressively modified by ensemble dephasing.

At low fields, $B/\Delta_B\lesssim1$, the fluctuating Overhauser field is comparable in magnitude to the applied field. Both the magnitude and orientation of the total effective field therefore vary significantly between nuclear-spin realizations, resulting in strong inhomogeneous spin dephasing. As $B/\Delta_B$ increases above unity, the applied field increasingly defines the spin-precession axis and the relative influence of transverse nuclear-field fluctuations is reduced. The oscillatory field dependence of $V_0$ consequently approaches the coherent-precession limit predicted by the Merkulov model.

At still larger fields, the measured visibility becomes systematically smaller than the quasi-static Merkulov prediction. One contribution arises from the finite radiative lifetime of the QD: as the Larmor frequency increases, the random photon-emission time associated with spontaneous decay translates into an increasing uncertainty in the spin-precession phase at emission, producing an additional emission-induced dephasing mechanism~\cite{Ramesh_Annoni_Margaria_Fioretto_Pishchagin_Morassi_Lemaître_Doty_Senellart_Lanco_etal._2025, Huet_Ramesh_Wein_Coste_Hilaire_Somaschi_Morassi_Lemaître_Sagnes_Doty_etal._2025}. Other deviations from the quasi-static description may also arise as the low-field approximation underlying the optical selection rules becomes less accurate.

For delays longer than the characteristic Overhauser-induced dephasing timescale, the coherent oscillations are strongly averaged and the field dependence evolves toward the usual Hanle-type response~\cite{Bracker_Stinaff_Gammon_Ware_Tischler_Shabaev_Efros_Park_Gershoni_Korenev_etal._2005}. In this regime, the field controls how efficiently the initial $z$-spin information is preserved between the two emission times. The resulting change in $\left|\langle\hat{S}_z(T)\hat{S}_z(0)\rangle\right|$ is directly accessed by $V_0$. For $T=6.9~\mathrm{ns}$, shown in panel~\textbf{(b)}, the measured field dependence displays the characteristic Hanle profile over a magnetic-field scale set by $\Delta_B$.

\begin{figure}[H]
\centering
\includegraphics[scale=0.55]{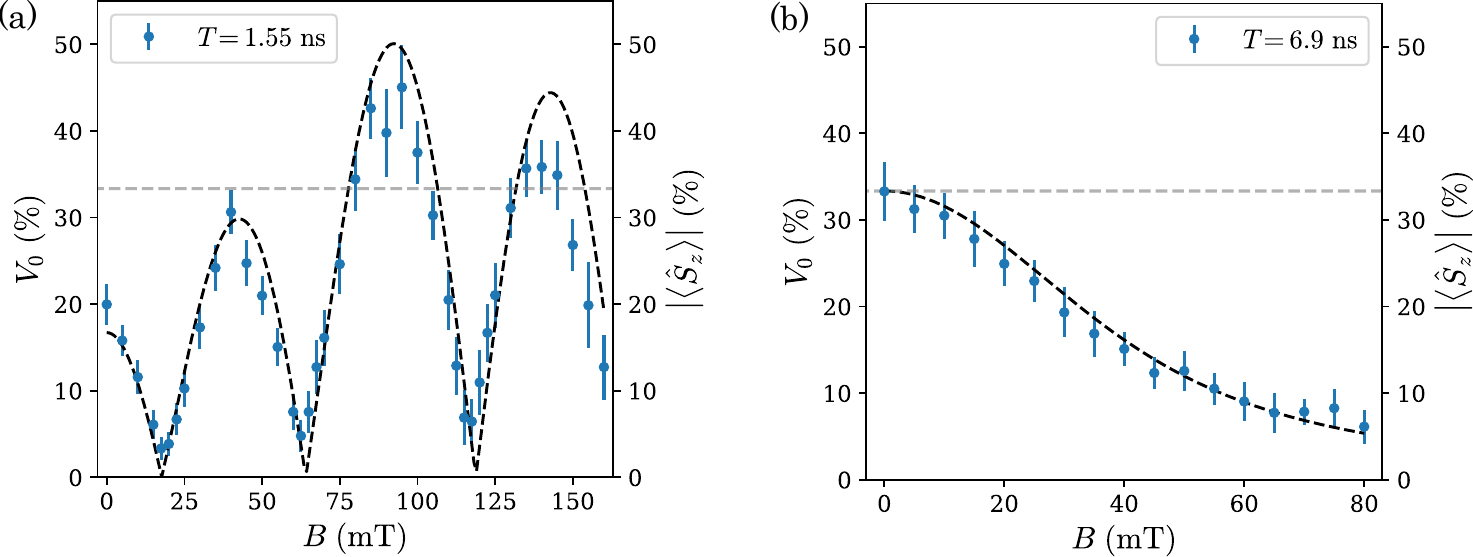}
\caption{\textbf{Spin-phase information transfer for B-field controlled gate.}Vacuum--one-photon interference visibility $V_0$ measured while varying the applied magnetic field $B$. The two panels correspond to delays $T=1.55~\mathrm{ns}$ and $6.9~\mathrm{ns}$, probing spin dynamics within and beyond the characteristic Overhauser-induced dephasing timescale, respectively. The dashed lines show the theoretical two-time spin correlation $\left|\langle\hat{S}_z(T)\hat{S}_z(0)\rangle\right|$ obtained from the Merkulov model~\cite{Merkulov_Efros_Rosen_2002}. Deviations at large magnetic fields arise from effects beyond the quasi-static model, including emission-induced spin dephasing and departures from the low-field optical-selection-rule approximation.}
\label{fig_hanle}
\end{figure}

\paragraph{Spin characteristics}
The same Merkulov model used to describe the temporal and magnetic-field dependence of $V_0$ also provides quantitative information on the QD spin environment. Fitting the dataset shown in Fig.~\ref{fig_merkulov} yields an Overhauser-induced dephasing timescale $T_{\mathrm O}=0.88\pm0.01\,\mathrm{ns}$ and an effective nuclear-field dispersion $\Delta_B=27.0\pm0.8\,\mathrm{mT}$. The comparison between the measured visibility and the modeled spin correlation gives a spin-phase information transfer fidelity $F=95.8\pm2.1\%$, increasing to $F=97.7\pm1.3\%$ when only data with $T>1\,\mathrm{ns}$ are retained, consistently with the additional interference effects discussed above at short delays.

Within the convention used for the Merkulov model, the electron Landé factor is related to these two fitted parameters through
\begin{equation}
|g_e|=\frac{\hbar}{\Delta_B\mu_B T_{\mathrm O}},
\end{equation}
giving $|g_e|=0.48\pm0.02$. This provides an independent spin parameter extracted directly from the time-dependent vacuum--one-photon interference measurements.

\begin{figure}[H]
\centering
\includegraphics[scale=0.8]{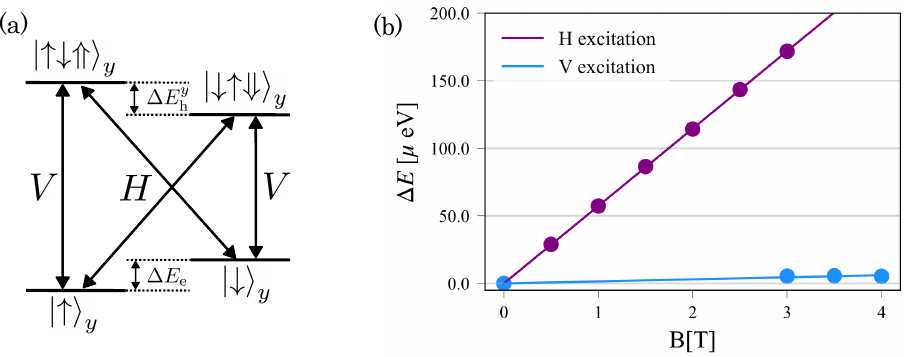}
\caption{\textbf{QD spin characterization.} \textbf{(a)} Optical selection rules of the trion under a strong Voigt magnetic field applied along $y$. \textbf{(b)} Spectral line splitting extracted from magnetoluminescence, showing the Zeeman splitting of the optical transitions, used to extract the electron $g_e$ factor.}
\label{fig_spincharacterization} 
\end{figure}

\paragraph{Comparison with continuous-wave spin measurements}
To verify that the spin properties extracted from the phase-sensitive photonic measurements are consistent with conventional spin characterization, we compare them with independent polarization-resolved continuous-wave measurements under an applied Voigt magnetic field. The corresponding optical selection rules are shown in Fig.~\ref{fig_spincharacterization}\textbf{(a)}. In the strong-field regime, the electron eigenstates are $\ket{\ua_y}$ and $\ket{\da_y}$, separated by $\Delta E_e=|g_e|\mu_B B$, while the two trion states are split by the hole Zeeman energy $\Delta E_h=|g_h|\mu_B B$. The four allowed optical transitions form a double-$\Lambda$ system, with two orthogonal linear polarizations $H$ and $V$.

For the two $V$-polarized transitions, the electron and hole Zeeman shifts enter with opposite signs, whereas they add for the $H$-polarized transitions. The energy splittings between the two optical transitions of a given polarization are therefore
\begin{align}
    \Delta E^V
    &= \hbar\left|\omega_\ua^V-\omega_\da^V\right|
    = \left|g_e-g_h\right|\mu_B B,\\
    \Delta E^H
    &= \hbar\left|\omega_\ua^H-\omega_\da^H\right|
    = \left|g_e+g_h\right|\mu_B B.
\end{align}
Here, $\Delta E^{H,V}$ correspond to the energy separations between the two $H$- or $V$-polarized resonances plotted in Fig.~\ref{fig_spincharacterization}\textbf{(b)}. Their linear dependence on the magnetic field gives the slopes $ \alpha_V=\left|g_e-g_h\right|\mu_B, \alpha_H=\left|g_e+g_h\right|\mu_B$. Assuming the level ordering shown in Fig.~\ref{fig_spincharacterization}\textbf{(a)}, these two measurements allow the electron and hole $g$-factors to be independently extracted as $|g_e|=\frac{\alpha_H+\alpha_V}{2\mu_B}$, $ |g_h|=\frac{\alpha_H-\alpha_V}{2\mu_B}$. We extract $|g_e|=0.5$ and $|g_h|=0.48$, in good agreement with $|g_e|=0.48\pm0.02$ extracted from the phase-sensitive measurement. This agreement supports the interpretation of the vacuum--one-photon interference visibility as a direct probe of the underlying QD spin dynamics.

\end{document}